\documentclass[a4paper,twocolumn]{revtex4}
\pdfoutput=1
\usepackage{graphicx}
\usepackage{slashed}
\usepackage{color}
\usepackage{xspace}
\usepackage[utf8]{inputenc}
\usepackage{amssymb}

\newcommand{\be}{\begin{equation}}
\newcommand{\ee}{\end{equation}}
\newcommand{\bea}{\begin{eqnarray}}
\newcommand{\eea}{\end{eqnarray}}

\newcommand{\unit}[1]{\ensuremath{\mathrm{\,#1}}\xspace}

\newcommand{\cm}{\unit{cm}}

\newcommand{\second}{\unit{s}}

\newcommand{\cmcubes}{\ensuremath{\cm^{3}\second^{-1}}\xspace}

\newcommand{\sigmavnn}{(\sigma v)_{\nu\bar\nu}}

\begin{document}
\title{Gamma-ray Limits on Neutrino Lines}
\author{Farinaldo S. Queiroz}
\email{farinaldo.queiroz@mpi-hd.mpg.de}
\author{Carlos E. Yaguna}\email{carlos.yaguna@mpi-hd.mpg.de}
\affiliation{\small \it Max-Planck-Institut f\"ur Kernphysik,%
 Saupfercheckweg 1, 69117 Heidelberg, Germany}
\author{Christoph Weniger}\email{c.weniger@uva.nl}
\affiliation{\small \it GRAPPA Institute, University of Amsterdam,
 Science Park 904, 1090 GL Amsterdam, Netherland}

\begin{abstract}
Monochromatic neutrinos from dark matter annihilations ($\chi\chi\to \nu\bar\nu$) are always produced in association with a gamma-ray spectrum  generated by  electroweak bremsstrahlung. Consequently, these \emph{neutrino lines} can  be searched for not only with neutrino detectors but also indirectly with gamma-ray telescopes. Here, we derive limits  on the dark matter annihilation cross section  into neutrinos based on  recent Fermi-LAT and HESS data. We find that, for dark matter masses above 200 GeV,  gamma-ray data  actually set the most stringent constraints on neutrino lines from dark matter annihilation and, therefore, an upper bound on the dark matter total annihilation cross section. In addition, we point out that gamma-ray telescopes, unlike neutrino detectors, have the potential to  distinguish the flavor of the final state  neutrino. Our results indicate that we have already entered into a new era where  gamma-ray telescopes are more sensitive than neutrino detectors to neutrino lines from dark matter annihilation.    
\end{abstract}

\maketitle

\section{Introduction}
Within the WIMP paradigm, the dark matter self-annihilation cross section is one of the fundamental properties of the dark matter particle. This cross section not only determines the relic density via thermal freeze-out in the early Universe, but also the expected indirect detection signatures of the dark matter. They typically consists of fluxes of gamma rays, neutrinos,  and antimatter, produced by dark matter annihilation in astrophysical objects, that could be visible over the expected background (for recent reviews see \cite{Cirelli:2015gux}, and in context of gamma rays \cite{Bringmann:2012ez}). Several experiments have been looking for these indirect detection signals, and so far no evidence of dark matter annihilation has been found in any channel. As a result, strong constraints on the dark matter annihilation cross section into different final states have been derived. Among these final states, the neutrinos are special because, being the least detectable of the SM particles, they provide an upper bound on the dark matter total annihilation cross section \cite{Beacom:2006tt}. Thus, the annihilation cross section into neutrinos, $\sigmavnn$, plays a key role in dark matter phenomenology. 

A direct way of constraining $\sigmavnn$ is by using data from neutrino detectors, as originally done in \cite{Beacom:2006tt,Yuksel:2007ac}. Based on atmospheric neutrino data from Fr\'ejus \cite{Daum:1994bf}, AMANDA \cite{Ahrens:2002gq}, and Super-Kamiokande \cite{Ashie:2005ik} detectors, and considering the cosmic and the halo signals, they set a bound on  $\sigmavnn$ of order $10^{-21}-10^{-22}\cmcubes$  over a wide range of dark matter masses. It was realized early on, though, that this annihilation cross section could also be indirectly constrained with gamma-ray data \cite{Kachelriess:2007aj,Bell:2008ey}. The final state neutrinos may, in fact, radiate W/Z bosons, which decay and eventually produce gamma rays at lower energies. Comparing the diffuse photon flux from the dark matter halo of our Galaxy against data from the EGRET satellite \cite{Hunger:1997we,Sreekumar:1997un}, the authors of \cite{Kachelriess:2007aj} derived, for dark matter masses between 100 GeV and 10 TeV,  a bound of order $10^{-19}-10^{-21}\cmcubes$ on $\sigmavnn$ --weaker than (or comparable to) the direct limit from neutrino detectors at that time (circa 2007). But, where do we stand today? Are  the direct bounds from neutrino detectors still stronger than the indirect limits from gamma-rays? 

Many things have changed in the meantime. Regarding neutrino detectors, we now have constraints from Super-Kamiokande and IceCube, which have improved by more than one order of magnitude the limits on $\sigmavnn$. Regarding gamma-ray telescopes, Fermi-LAT and H.E.S.S. have been exploring the gamma-ray sky with unprecedented sensitivity. Data from these telescopes have already been used to strongly constrain the dark matter annihilation cross section into different final states. Moreover, gamma-rays typically provide,  among the different indirect detection channels,  the most stringent and robust bounds on the dark matter annihilation cross section. It may well be, therefore, that gamma-ray telescopes have already catch up with neutrino detectors in the search for neutrino lines.  

In this paper we derive Fermi-LAT and H.E.S.S. limits on neutrino lines from dark matter annihilation, as well as projected sensitivities for CTA (Cherenkov Telescope Array). After comparing our limits  against recent bounds from Super-Kamiokande and IceCube, we find that, for dark matter masses above 200 GeV,  gamma-ray data set the most stringent constraint on neutrino lines and, therefore, the upper bound on the dark matter total annihilation cross section into standard model particles.  Our results further suggest that a new era has begun where  gamma-ray telescopes are more sensitive  to neutrino lines from dark matter annihilation than neutrino detectors.

Keep in mind that this finding does not undervalue the role of neutrino telescopes in probing neutrino lines from dark matter annihilation, since they can truly determine whether the observed signal is in fact a neutrino line, differently from gamma-rays telescopes. Moreover, only neutrino telescopes can probe low dark matter masses (below $200$~GeV) where electroweak corrections are irrelevant.

In what follows, we explain how one can a gamma-ray emission in induced from neutrino lines from dark matter
annihilation.

\section{Gamma rays from  neutrino lines}
\begin{figure}[tb]
\begin{tabular}{ccc}
\includegraphics[scale=0.4,trim= 0 0 1.5cm 0,clip]{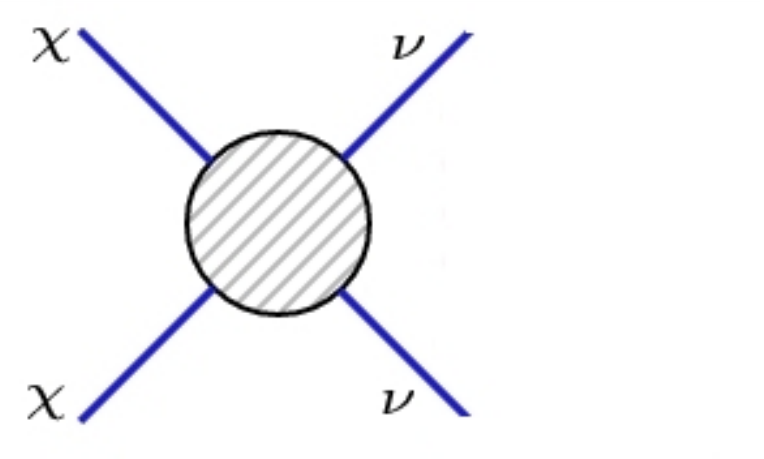} & \includegraphics[scale=0.4,trim= 0 0 1.0cm 0,clip]{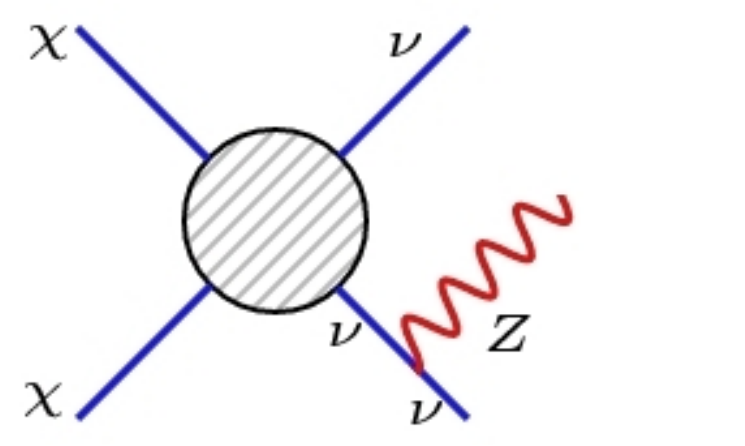}  & \includegraphics[scale=0.4,trim= 0 0 1.5cm 0,clip]{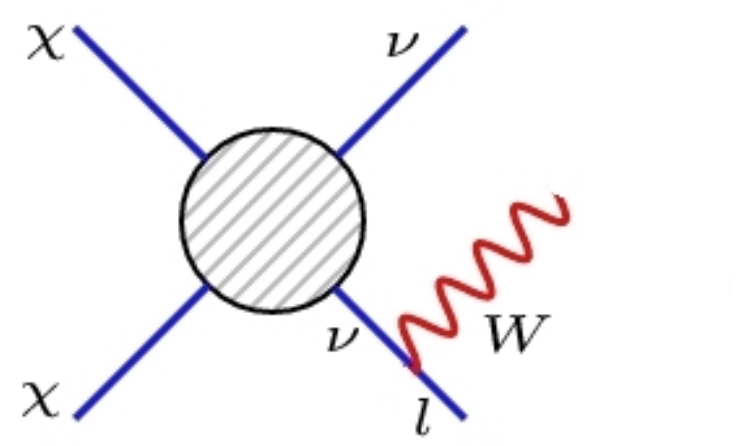}
\end{tabular}
\caption{An illustration of some weak corrections to dark matter annihilation into neutrinos. The final state neutrinos can emit gauge bosons that will eventually decay producing gamma rays. \label{fig:diagrams}} 
\end{figure} 

Weak corrections may play a very important role in dark matter indirect detection \cite{Kachelriess:2007aj,Bell:2008ey,Kachelriess:2009zy,Ciafaloni:2010ti,Bell:2011if,Bringmann:2013oja,Bringmann:2011ye,Weniger:2011ik}. The reason is that, for dark matter masses above the weak scale, soft electroweak gauge bosons can be copiously radiated from highly energetic final states, significantly modifying the  spectrum of annihilation products. These modifications can be  essentially of two different types. On the one hand, they may change the low energy part of the spectrum, as they tend to convert a small number of highly energetic particles into a large number of low energy ones. On the other hand, they may give rise to the appearance of  new channels in the final states that would be absent if such corrections were neglected. This second effect is the relevant one for dark matter annihilation into neutrinos. When  weak corrections are neglected, the gamma-ray flux from the annihilation  of dark matter particles into monochromatic neutrinos ($\chi\chi\to \nu\bar\nu$) vanishes.  But once the weak corrections are taken into account, the final state neutrinos may emit $W$ and $Z$ gauge bosons which would in turn decay and eventually produce a continuous gamma-ray spectrum at lower energies --see figure \ref{fig:diagrams}. Thus, weak corrections lead to the existence of a  gamma-ray spectrum associated with neutrino lines.

More generally, weak corrections imply that  whenever the dark matter particle annihilates or decays into SM fields, \emph{all} stable particles ($\gamma$, $\nu$, $e^+$, $\bar p$) will be present in the prompt spectrum at some degree. Neutrino lines, for instance, will not only produce gamma rays but also antiprotons and positrons, and a continuum spectrum of neutrinos.

When  electroweak corrections are included, the predicted gamma-ray spectrum from neutrino lines contains logarithmically-enhanced terms of the form $\alpha_2\ln^2M_{DM}^2/M_W^2$ and $\alpha_2\ln M_{DM}^2/M_W^2$, where $M_{DM}$ is the dark matter mass.  Thus, they are particularly relevant at high dark matter masses. These corrections are already  implemented into the PPPC code \cite{Cirelli:2010xx}, on which we will rely in the following to obtain the gamma-ray spectra from dark matter annihilations. Let us  emphasize that the gamma-ray spectrum associated with neutrino lines is model-independent \cite{Ciafaloni:2010ti}. The only assumptions used in its derivation are that the SM describes the physics up to the scale $M_{DM}$, and that, from then on,  the SM can be extended by physics that preserves the SM gauge invariance.

\begin{figure}[tb]
\includegraphics[scale=0.30]{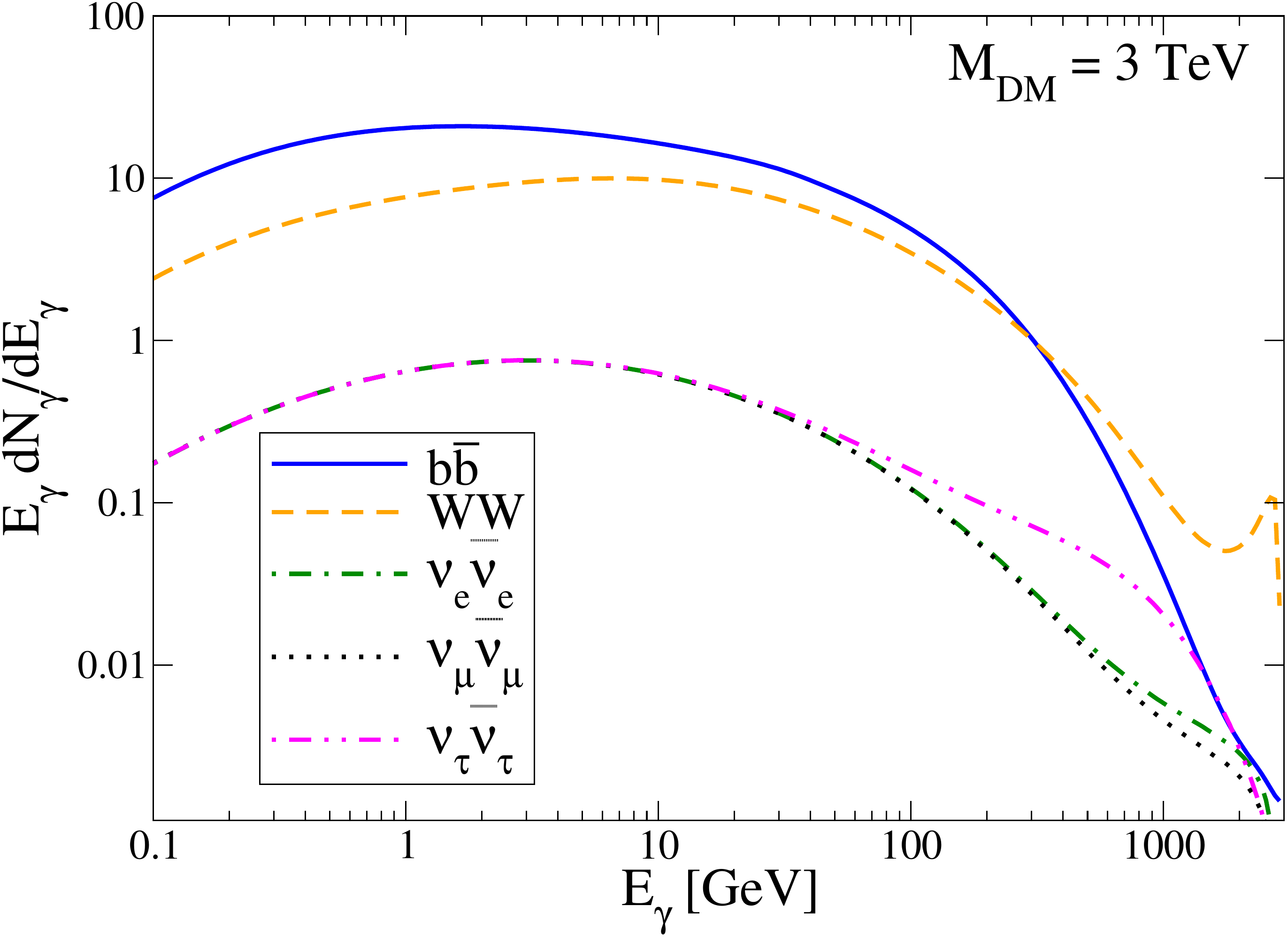}
\caption{A comparison between the gamma ray fluxes produced by dark matter annihilation into different final states: $b\bar b$, $WW$, $\nu_\alpha\bar\nu_\alpha$. The  dark matter mass set to  3 TeV. Notice that annihilation into neutrinos gives rise to a gamma ray spectrum which is suppressed by about one  order of magnitude with respect to conventional channels.  \label{fig:fluxes}} 
\end{figure} 
\medskip
The gamma-ray spectrum produced by dark matter annihilation into neutrinos is shown in figure \ref{fig:fluxes} for a dark matter mass of $3$ TeV. For comparison, the spectra from the standard annihilation channels $b\bar b$ and $W^+W^-$ are also displayed.  Notice that the spectral shape is rather similar for all the final states. We can see from the figure that dark matter annihilation into neutrinos gives rise to a gamma-ray spectrum which is suppressed by about one order of magnitude with respect to that for the  $W^+W^-$ final state. Thus,  for a dark matter mass of $3$ TeV, gamma-ray data should set a limit on  $\sigmavnn$  of order $10^{-23}~\mathrm{cm^3s^{-1}}$, or about a factor ten weaker than the current Fermi-LAT limit for annihilation into $W^+W^-$ \cite{Ackermann:2015zua}. This factor will tend to decrease toward higher dark matter masses as the electroweak corrections become more important. Notice also that the  spectrum for the  $\nu_\tau$ final state diverges   from that for  $\nu_{e},\nu_\mu$ at high energies. The reason for this difference is the tau-lepton that necessarily accompanies the W boson emitted by $\nu_{\tau}$ --see figure \ref{fig:diagrams}.  At high energy, a tau-lepton generates a harder gamma-ray spectrum than an electron or a muon  \cite{Cirelli:2010xx}, accounting for the effect observed in the figure. This effect implies that, unlike neutrino detectors, gamma-ray telescopes may potentially  distinguish among the different neutrino flavors into which the dark matter annihilates.
\section{Constraints on \boldmath$\sigmavnn$}
\begin{figure}[tb]
\includegraphics[scale=0.3]{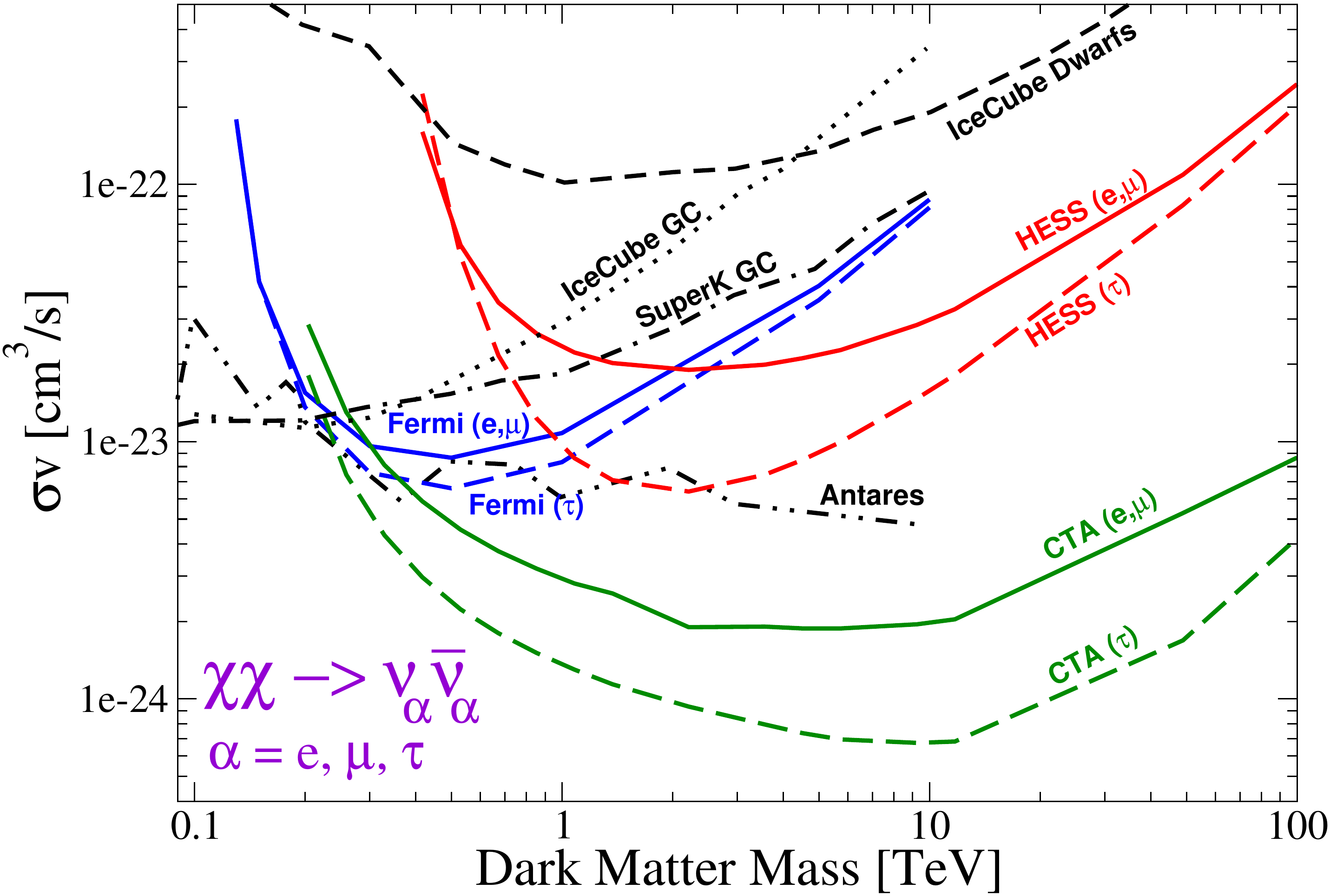} 
\caption{Limits on neutrino lines from dark matter annihilation. The colored lines show the gamma-ray bounds derived in this work using  Fermi-LAT data on dwarfs (blue) and H.E.S.S. data from the GCH (red). The black lines show current constraints from neutrino detectors. Notice that, for dark matter masses above 200 GeV, the gamma-ray limits provide the most stringent constraint on neutrino lines. The green lines show the expected CTA sensitivity. Solid lines indicate the limits for a $\nu_{e,\mu}$ final state while dashed lines correspond to $\nu_\tau$.  \label{fig:ann}. {\it Note}: Posterior to publication we became aware of the recent ANTARES limit from Galactic Center observation \cite{Adrian-Martinez:2015wey}.} 
\end{figure}

Monochromatic neutrinos from dark matter annihilations, $\chi\chi\to \nu_{\alpha}\bar\nu_{\alpha}$, have been studied before in several contexts --see e.g. \cite{Barger:2001ur,Barger:2007hj,Lindner:2010rr,Farzan:2011ck,Arina:2015zoa}. They give rise to a differential photon flux,  from a given angular direction $\Delta\Omega$, that can be expressed as
\begin{equation}
\frac{d\Phi_\gamma(\Delta \Omega)}{dE}(E_\gamma)=\frac{1}{4\pi}\frac{\sigmavnn}{2 M_{DM}^2}\frac{dN_\gamma}{dE_{\gamma}}\cdot J_{ann}\;,
\end{equation}
where $J_{ann}$ is the annihilation J-factor, $M_{DM}$ is the dark matter mass, and $\frac{dN_\gamma}{dE_{\gamma}}$ is the differential $\gamma$-ray yield per annihilation into the $\nu_{\alpha}\bar\nu_{\alpha}$ final state. The idea then  is to use gamma-ray data to constrain regions in the ($M_{DM}$, $\sigmavnn$) plane for dark matter annihilation into a given neutrino flavor. In our analysis we derive  limits from Fermi-LAT and H.E.S.S. data, and obtain projected sensitivities for CTA. Our results are shown in figure \ref{fig:ann}.

The Fermi-LAT team provided \cite{Fermidata} tabulated likelihood functions for the pass8 dwarf spheroidal analysis \cite{Ackermann:2015zua}, which we have adopted in the present work to calculate upper limits (at 95\% CL) on $\sigmavnn$\footnote{A convenient wrapper for all likelihoods used in this work will be publicly available soon (gamLike, C. Weniger et al., 2016)}. These limits are displayed, for dark matter masses between $100$ GeV and $10$ TeV, as  solid ($\nu_e,\nu_\mu$) and dashed ($\nu_\tau$) blue lines in figure \ref{fig:ann}. As expected from the spectrum, the constraint is slightly stronger for annihilation into the $\nu_\tau$ flavor. This limit from dwarfs on $\sigmavnn$ reaches a minimum of $6\times 10^{-24}\mathrm{cm^3s^{-1}}$ for dark matter masses of about $500$ GeV. It decreases toward lower and higher masses reaching $10^{-23}\mathrm{cm^3s^{-1}}$ for dark matter masses of about $200$ GeV and $1.3$ TeV. At the upper end of the dark matter mass, for $M_{DM}=10$ TeV, the limit becomes of order $8\times10^{-23}\mathrm{cm^3s^{-1}}$.

The H.E.S.S. limits we derive  are based on the search from the Galactic  Center halo reported in \cite{Abramowski:2011hc, Viana:2012wka}. They assume a NFW profile but the result is very similar for a Einasto profile. In figure \ref{fig:ann} these limits are displayed as solid and dashed red lines, and they extend to dark matter masses of $100$ TeV. Notice that they are stronger than the Fermi-LAT limits for dark matter masses above $1$ TeV for the $\nu_\tau$ final state and above $2$ TeV for $\nu_{e,\mu}$. At $M_{DM}=10$ TeV, for instance, the H.E.S.S. bounds are about a factor $3$ ($\nu_{e,\mu}$) and $5$ ($\nu_\tau$) stronger than the Fermi-LAT ones. It is also clear from this figure that, being sensitive to more energetic gamma rays, H.E.S.S. can better distinguish the gamma-ray spectrum from different neutrino flavors.  At $2$ TeV, for example, the difference between the $\nu_\tau$ and the $\nu_{e,\mu}$ lines amounts to about a factor $3$. The H.E.S.S. collaboration recently presented new preliminary results \cite{Lefranc:2015vza} updating those in \cite{Abramowski:2011hc}. By using the whole H.E.S.S. I dataset (2004-2014) and a novel analysis technique, they were able to significantly improve the constraints on the dark matter annihilation cross section, particularly at high masses.  At 10 TeV, for instance, the new reported limits are one order of magnitude stronger than the previous one (for the $b\bar b$ final state). Since these data are not yet publicly available, we could not use them in our analysis, but it is clear that  the H.E.S.S. limits on neutrino lines will also become much more stringent once such data are taken into account.

In addition to our own limits from Fermi-LAT and H.E.S.S., we also displayed in figure \ref{fig:ann} current limits on the annihilation cross section into neutrinos from IceCube and Super-Kamiokande (see \cite{Covi:2009xn, Esmaili:2012us, Aisati:2015vma} for  searches focused on dark matter decay). The  dashed black line shows preliminary results, recently obtained by the IceCube collaboration, from the first search for dark matter annihilations in dwarf galaxies with the complete IceCube detector \cite{Aartsen:2015bwa}. They are based on a stacked analysis of the  5 dwarf galaxies with the largest J-factors. As can be seen in the figure, the resulting limit lies above $10^{-22}\mathrm{cm^3s^{-1}}$ over the entire mass range and is significantly weaker than the Fermi-LAT limits, which were also derived from the same targets. In fact, our gamma-ray limits improve this IceCube limit from dwarfs by more than one order of magnitude for some dark matter masses. Last year, the IceCube collaboration  published the results from a search for dark matter annihilation in the Galactic Center\cite{Aartsen:2015xej}. Their upper limits on $\sigmavnn$ are shown, for the NFW profile, as as dotted black line. The gamma-ray limit we derived is stronger than this limit for dark matter masses above $200$ GeV.  Specifically, the Fermi-LAT and H.E.S.S. limits we obtained are more stringent than the IceCube (GC) one by up to  factors of $4$ and $20$, respectively. The Super-Kamiokande collaboration has also recently reported preliminary results regarding the search for neutrinos from dark matter annihilation in the Galactic Center \cite{Frankiewicz:2015zma}.  Their resulting limit is shown as as a dash-dotted black line in figure \ref{fig:ann}. The gamma-ray limit from Fermi-LAT turns out to be more constraining for dark matter masses above $200$ GeV while the H.E.S.S. limits are so for dark matter masses above $700$ GeV ($\nu_{\tau}$) and $1.2$ TeV ($\nu_{e,\mu}$). Summarizing, we see that somewhat surprisingly gamma rays currently provide the most stringent constraints on neutrino lines from dark matter annihilation. 

As a corollary, the Fermi-LAT and H.E.S.S. limits we derived also provide a new upper bound on the total dark matter annihilation cross section. That is, provided that dark matter annihilates into SM two-body final states, its total annihilation rate, $(\sigma v)_{tot}$, necessarily satisfies $(\sigma v)_{tot}<\sigmavnn$ \cite{Beacom:2006tt}, with $\sigmavnn$ constrained according to figure \ref{fig:ann}.

Finally, in figure \ref{fig:ann} we also display, as green lines, the expected CTA sensitivity to neutrino lines from dark matter annihilation in the Galactic Center. These limits were obtained following the analysis in \cite{Silverwood:2014yza}. It assumes 100 hours of observations,  $1\%$ systematics and a Einasto profile. As can be observed in the figure, CTA will significantly improve the bounds on neutrino lines, particularly at high dark matter masses. It is apparent from this figure that  neutrino detectors have already lost the battle against gamma-ray telescopes in the search for neutrino lines from dark matter annihilation. We checked that for decaying dark matter instead, neutrino telescopes provide better limits than Fermi-LAT in the entire dark matter mass range in Fig.\ref{fig:ann}. We could not derive H.E.S.S sensitivity in this case because their results are based on the morphology of both signal and background contributions. 

\section{Conclusions}
We derived gamma-ray limits  on neutrino lines from dark matter annihilation. Our starting point was the fact that monochromatic neutrinos --neutrino lines-- from dark matter annihilation are always produced, via electroweak corrections, in association with a model-independent gamma-ray spectrum at lower energies.  Thus, it is feasible to indirectly search for neutrino lines using gamma-ray telescopes. We used Fermi-LAT data from a stack of dwarf galaxies and H.E.S.S. data from the Galactic center halo region to constrain the dark matter annihilation cross section  into neutrinos. We found that current gamma-ray limits  on neutrino lines from dark matter annihilation  are more stringent, for dark matter masses above $200$ GeV, than those derived from neutrino detectors such as IceCube and Super-Kamiokande. These limits constitute a new upper bound on the total dark matter annihilation cross section. In addition,  we  showed that the CTA has the potential to significantly improve these limits in the near future. Our results indicate that we have already entered into a new era where  gamma-ray telescopes are more sensitive than neutrino detectors to neutrino lines from dark matter annihilation.   

Note that our results do not undervalue the role of neutrino telescopes in probing neutrino lines from dark matter annihilation since only them can truly discriminate a neutrino line, differently from gamma-rays telescopes. Additionally, neutrino telescopes are indeed the most sensitive to low dark matter masses, below $200$~GeV, where electroweak corrections are turned off.

\section*{Acknowledgements}
We warmly acknowledge discussions with Alexei Smirnov. We thank John Beacom for reading the manuscript as well as Torsten Bringmann, Alessandro Strumia, Dan Hooper, Stefano Profumo and Pavel Fileviez Perez for comments. CY is supported by the Max Planck Society in the project MANITOP. CW acknowledges funding from an NWO Vidi grant.

\bibliography{darkmatter}

\end{document}